\documentclass{aa}  

\usepackage[dvipsnames]{xcolor}
\usepackage{graphicx}
\usepackage{mhchem}
\usepackage[normalem]{ulem}
\usepackage{color,soul}
\usepackage{lscape}
\usepackage{balance}
\usepackage{txfonts}
\usepackage{mathtools}
\usepackage{multirow}
\usepackage{booktabs}
\newcounter{chem}
\newcommand{\curly}{\mathrel{\leadsto}}
\newcounter{temp}

\begin{document} 
   \title{Hydroxylamine in Astrophysical Ices: IR Spectra and Cosmic Ray-Induced Radiolytic Chemistry}
   
\author{Belén~Maté\inst{1}
          \and
          Ramón~J.~Peláez\inst{1}
          \and
          Germán~Molpeceres\inst{2}
          \and
          Richárd~Rácz\inst{3}
          \and
          Duncan~V.~Mifsud\inst{3}
          \and
          Juan~Ortigoso\inst{1}
          \and
          Víctor~M.~Rivilla\inst{4}
          \and
          Gergő~Lakatos\inst{3,5,6}
          \and
          Béla~Sulik\inst{3}
          \and
          Péter~Herczku\inst{3}
          \and
          Sergio~Ioppolo\inst{7}
          \and
          Sándor~Biri\inst{3}
          \and
          Zoltán~Juhász\inst{3}
          }

\institute{Instituto de Estructura de la Materia, IEM-CSIC, Calle Serrano 121, 28006 Madrid, Spain\\
    \email{belen.mate@csic.es}
    \and
    Instituto de Fisica Fundamental, IFF-CSIC, Calle Serrano 121, 28006 Madrid, Spain
    \and
    HUN-REN Institute for Nuclear Research (Atomki), Bem Tér 18/C, Debrecen H-4026, Hungary
    \and
    Centro de Astrobiología (CAB), CSIC-INTA, Carretera de Ajalvir Km. 4, Torrejón de Ardoz, 28850 Madrid, Spain
    \and
    Institute of Chemistry, University of Debrecen, Egyetem Tér 1, Debrecen H-4032, Hungary
    \and
    Doctoral School of Chemistry, University of Debrecen, Egyetem Tér 1, Debrecen H-4032, Hungary
    \and
    Centre for Interstellar Catalysis (InterCat), Department of Physics and Astronomy, Aarhus University, Aarhus DK-8000, Denmark
    }

\date{\today}

\abstract
{Gas-phase hydroxylamine (NH$_2$OH) has recently been detected within dense clouds in the interstellar medium. However, it is also likely present within interstellar ices, as well as on the icy surfaces of outer Solar System bodies, where it may react to form more complex prebiotic molecules such as amino acids.}
{In this work, we aimed to provide IR spectra of NH$_2$OH in astrophysical ice analogues that will help in the search for this molecule in various astrophysical environments. Furthermore, we aimed to provide quantitative information on the stability of NH$_2$OH upon exposure to ionizing radiation analogous to cosmic rays, as well as on the ensuing chemistry and potential formation of complex prebiotic molecules.}
{Ices composed of NH$_2$OH, H$_2$O, and CO were prepared by vapor deposition and IR spectra were acquired between $4000-500$~cm$^{-1}$ ($2.5-20$~$\mu$m) prior and during irradiation using 15~keV protons.}
{Our spectroscopic characterizations have determined that NH$_2$OH ices deposited at $10-20$~K adopt an amorphous structure, which begins to crystallize upon warming to temperatures greater than 150~K. In interstellar ice analogues, the most prominent IR absorption band of NH$_2$OH is that at about 1188~cm$^{-1}$, which may be a good candidate to use in searches of this species in icy space environments. Calculated effective destruction cross-sections and \textit{G}-values for the NH$_2$OH-rich ices studied show that NH$_2$OH is rapidly destroyed upon exposure to ionizing radiation (more rapidly than a number of previously studied organic molecules), and that this destruction is slightly enhanced when it is mixed with other icy species. The irradiation of a NH$_2$OH:H$_2$O:CO ternary ice mixture leads to a rich chemistry that includes the formation of simple inorganic molecules such as NH$_3$, CO$_2$, OCN$^-$, and H$_2$O$_2$, as well as ammonium salts and, possibly, complex organic molecules relevant to life such as formamide, formic acid, urea, and glycine.}
   {}

\keywords{astrochemistry -- astrobiology -- radiation chemistry -- infrared spectroscopy -- hydroxylamine -- ices}

\maketitle

\section{Introduction}
\label{sec:introduction}
The rich diversity of molecules known to populate the interstellar medium \citep{mcguire20222021} has spurred astrochemists to search for so-called \textit{prebiotic molecules} whose chemical processing could conceivably lead to the formation of the fundamental building blocks of life-relevant compounds (i.e., amino acids, nucleobases, sugars, and lipids). Indeed, such searches have recently met with considerable success, with a number of prebiotic molecules such as urea (H$_2$NCONH$_2$), \textit{Z}-ethenediol (HOC$_2$H$_4$OH), and ethanolamine (HOCH$_2$CH$_2$NH$_2$) having been detected in dense molecular clouds over the past few years \citep{belloche2019re,rivilla2021discovery,rivilla2022precursors}.

One such molecule that has been detected recently in the gas-phase within dense molecular clouds is hydroxylamine (NH$_2$OH) \citep{rivilla2020prebiotic}. This molecule, a derivative of ammonia, is known to be a key intermediate during the synthesis of amino acids and nucleotides \citep{sakurai1984prebiotic,snow2007gas,becker2019unified,zhu2020production,xu2022isoxazole}. It is therefore perhaps unsurprising that hydroxylamine has been the focus of several past astrochemical studies. Initial searches for this molecule in interstellar regions proved unsuccessful \citep{mcguire2015cso,ligterink2018methanol}, until it was finally definitively detected by \citeauthor{rivilla2020prebiotic} in 2020. Laboratory studies have also considered viable methods for its formation and chemical evolution \citep{charnley2001gas,congiu2012no,fedoseev2016simultaneous,tsegaw2017formation,nguyen2019experimental}, while modeling efforts have focused on its relative abundance in the interstellar medium \citep{he2015formation,garrod2022formation,molpeceres2023processing}. All efforts thus far seem to suggest that hydroxylamine is primarily synthesized in the solid phase within interstellar icy grain mantles, before being released to the gas phase through desorption or sublimation. 

\citet{rivilla2020prebiotic} showed that gas-phase hydroxylamine in dense molecular clouds has a relatively low abundance of $2.1\times10^{-10}$ with respect to H$_2$. This observation is in line with predictions recently made by our chemical models \citep{molpeceres2023processing}, which have further predicted that the content of hydroxylamine within interstellar icy grain mantles should be enhanced to abundances of $10^{-7}-10^{-8}$ with respect to H$_2$. This would make hydroxylamine a comparatively abundant prebiotic molecule within ices. Given that hydroxylamine possesses both a hydroxyl ($-$OH) and an amine ($-$NH$_2$) group, as well as a relatively labile N$-$O bond in its structure, it is conceivable that its radiolytic processing in the solid phase within interstellar icy grains mantles by impinging cosmic rays could lead to the formation of more complex prebiotic species, such as the aforementioned urea and ethanolamine, as well as potentially glycolamide (H$_2$NCOCH$_2$OH) or glycine (H$_2$NCH$_2$COOH). 

As such, an accurate understanding of the radiolytic destruction (or stability) of hydroxylamine is imperative if its contribution to the formation of biologically relevant molecules in interstellar space is to be constrained. Although there exists a kinetic formalism for the inclusion of radiolysis in astrochemical models \citep{shingledecker2018general}, there is no theoretical framework to predict the likelihood of possible radiolytic reactions under astrophysical conditions. As such, laboratory experiments are critical to assess the cosmic ray-induced radiation chemistry occurring in interstellar ices and thus provide data which may be fed into astrochemical models. Indeed, there exists of wealth of studies that has demonstrated that astrophysical ice analogues, when irradiated by ion or electron beams mimicking cosmic rays, host the synthesis of various molecules of direct relevance to biology and the origins of life \citep{hudson2008amino,zhu2020production,kleimeier2021cyclopropenone,turner2021photoionization,marks2023complex,zhang2023formation,wang2024interstellar}.

In this present study, we considered astrophysical ice analogues rich in hydroxylamine prepared by vapor deposition onto a cooled surface. Both single-component ices, as well as binary mixtures with H$_2$O and ternary mixtures with H$_2$O and CO have been studied. IR absorption spectra of these ices have been acquired with the aim of generating spectroscopic data that may be of use to current high-resolution telescopes (e.g., the \textit{James Webb Space Telescope}) in the search for solid-phase hydroxylamine in the interstellar medium or on the surfaces of icy outer Solar System bodies. The neat hydroxylamine ices, together with the binary and ternary mixtures, were then subjected to irradiation using 15~keV protons as cosmic ray analogues with the aim of quantifying the radiolytic destruction of hydroxylamine under realistic astrophysical conditions and characterizing the chemical products of irradiation. Further details of the experimental apparatus and methodologies used in this study are given in Section~\ref{sec:methods}, and the results of our study are presented and discussed in Sections~\ref{sec:results} and~\ref{sec:discussion}, respectively. Finally, concluding remarks are given in Section~\ref{sec:conclusions}.

\section{Apparatus and Methodology}
\label{sec:methods}
Experiments were carried out at two laboratories: IR spectroscopic characterizations were performed at the Instituto de Estructura de la Materia of the Consejo Superior de Investigaciones Científicas (IEM-CSIC) in Madrid, Spain, while proton irradiation studies were carried out at the HUN-REN Institute for Nuclear Research (Atomki) in Debrecen, Hungary.

\subsection{Experiments Performed at IEM-CSIC}
The experimental set-up at IEM-CSIC was described in detail in previous publications (e.g., \citet{mate2024indene}). Briefly, the set-up consists of a high-vacuum chamber having a base pressure of $5\times10^{-8}$~mbar at room temperature. The chamber is equipped with a closed-cycle helium cryostat as well as a Fourier-transform IR spectrophotometer (Bruker Vertex 70) with an external MCT detector. Within the chamber is a 2~mm-thick KBr substrate held in thermal contact with the cold finger of the cryostat whose temperature can be controlled between $10-300$~K with an accuracy of 0.5~K.

Neat hydroxylamine ices were prepared by the deposition of the vapor generated by the thermal decomposition of the solid hydroxylammonium phosphate ((NH$_3$OH$^+$)$_3$PO$_4^{3-}$) salt, supplied by Santa Cruz Biotechnology. The salt was placed in a home-made sublimation oven located within the chamber and the cooled KBr deposition substrate was rotated to directly face the oven. The oven was subsequently warmed to 37~\textdegree{}C and, having equilibrated at this temperature, its shutter was opened to allow for the deposition of hydroxylamine onto the substrate surface. After several minutes of deposition, the oven shutter was closed and the deposition substrate was rotated to face the incident IR spectroscopic beam. Transmission absorption spectra were acquired as an average of 300 scans across a wavenumber range of $4000-500$~cm$^{-1}$ and at a resolution of 4~cm$^{-1}$.

\subsection{Experiments Peformed at HUN-REN Atomki}
Experiments at HUN-REN Atomki were carried out using the recently commissioned AQUILA set-up, which has been described in detail in the work of \citet{racz2024aquila}. The AQUILA consists of an ultrahigh-vacuum chamber operating at a base pressure of a few 10$^{-9}$~mbar which is installed as a permanent end-station to an electron cyclotron resonance ion source (ECRIS) that can supply a number of versatile ion beams able to simulate extra-terrestrial radiation \citep{biri1997new,biri2021atomki,racz2012molecular}. Within the center of the chamber is a 3~mm-thick ZnSe deposition substrate that is held in thermal contact with the cold finger of a closed-cycle helium cryostat, allowing the temperature of the substrate to be controlled between $20-300$~K with an accuracy of 0.2~K.

Similarly to the experiments conducted at IEM-CSIC, hydroxylamine ices could be prepared on the cooled ZnSe substrate through the thermal decomposition of the (NH$_3$OH$^+$)$_3$PO$_4^{3-}$ salt; this time using a commercial effusive evaporator (Createc OLED-40-10-WK-SHM). After rotating the deposition substrate to directly face the nozzle of the effusive evaporator, the salt was warmed to 37~\textdegree{}C and allowed a few minutes to equilibrate at this temperature. Subsequently, the shutter to the effusive evaporator was opened for a few minutes to allow for hydroxylamine vapor to condense onto the cooled deposition substrate.

To prepare binary and ternary ices containing hydroxylamine mixed with H$_2$O (or D$_2$O) and~/~or CO, vapors of these volatile species were introduced into the main chamber of the AQUILA through a separate dosing line during the deposition of the hydroxylamine so as to generate a co-deposited ice mixture. It is important to note that the geometry of the chamber as well as the radiation shield surrounding the cooled ZnSe deposition substrate prevent significant deposition of the volatile species onto the rear side of the substrate (see \citet{herczku2021ice} and \citet{racz2024aquila} for more details). 

Once an ice had been prepared, the deposition substrate was rotated to face an incident IR spectroscopic beam supplied by a Bruker Vertex 70v spectrophotometer (coupled to an external MCT detector) and a transmission absorption spectrum of the prepared ice was acquired as an average of 300 scans over a wavenumber range of $4000-650$~cm$^{-1}$ and at a resolution of 4~cm$^{-1}$. This spectrum could then be used to quantify the abundance of a certain species of molecule within the ice through measurement of its column density $N$ (molecules cm$^{-2}$), which is related to the integrated absorbance of a characteristic IR absorption band, $I$ (cm$^{-1}$), as follows:

\begin{equation}
\label{eq1}
 N=\textup{ln}(10)\:\frac{I}{A_{\nu}}
\end{equation}

\noindent where $A_\nu$ is the band strength constant associated with that particular band. 

Although \citet{luckhaus1997rovibrational} provided experimental and theoretical band strength constants for gas-phase hydroxylamine (quantifying $A_\nu=2.9\times10^{-18}$~cm molecule$^{-1}$ for the NH$_2$ wagging mode at 1115~cm$^{-1}$), to the best of our knowledge band strength constants for solid hydroxylamine at low temperatures have not yet been evaluated. Indeed, previous astrochemical studies have had to make assumptions regarding the band strength constants of the IR absorption features of solid-phase hydroxylamine. For example, \citet{tsegaw2017formation} made use of a band strength of $A_\nu=4.16\times10^{-18}$~cm molecule$^{-1}$ for the NOH bending mode at 1486~cm$^{-1}$, which they adopted from a theoretical calculation performed by \citet{saldyka2003photodecomposition} for the NH$_2$OH$-$CO dimer. On the other hand, \citet{jonusas2016possible} made use of a band strength constant of $A_\nu=1.2\times10^{-17}$~cm molecule$^{-1}$ for the $-$NH$_2$ wagging mode at 1194~cm$^{-1}$. This value was itself adopted by comparing against a similar vibrational mode (i.e., the umbrella mode at 1070~cm$^{-1}$) in the spectrum of solid NH$_3$ ice, for which the reported band strength ranges from $A_\nu=1.3\times10^{-17}$~cm molecule$^{-1}$ \citep{zanchet2013optical} to $A_\nu=1.7\times10^{-17}$~cm molecule$^{-1}$ \citep{hudson2022ammonia}. In this present work, we have experimentally estimated the band strength constant of the NH$_2$ wagging mode (which appears at about 1188~cm$^{-1}$ in our spectra) to be $A_\nu=9\times10^{-18}$~cm molecule$^{-1}$, with an expected uncertainty of 30\%. Details on how this value was estimated may be found in the appendix.

Having quantified the column density of hydroxylamine in this way, the thickness of a neat hydroxylamine ice, $h$ ($\mu$m), could be calculated as:

\begin{equation}
\label{eq2}
 h=10^4\:\frac{Nm}{\rho N_A}
\end{equation}

\noindent where $m$ is the molar mass of the molecule (g mol$^{-1}$), $\rho$ is the ice density (g cm$^{-3}$), and $N_A$ is the Avogadro constant. 

As is evident from Eq.~\ref{eq2}, calculating the thickness of the ice in this way requires knowledge of not only the column density of the deposited hydroxylamine, but also of its density. \citet{meyers1955crystal} reported a density of 1.4~g cm$^{-3}$ for a perfect crystal of hydroxylamine, but it appears that the density of the amorphous solid at temperatures relevant to astrochemistry has not yet been measured. We therefore adopt an estimated density of 1~g cm$^{-3}$ for the amorphous solid at 20~K, based on the typical lower density of vapor-deposited amorphous solids compared to the density of their corresponding crystalline structures as measured through crystallographic techniques. 

The use of Eq.~\ref{eq2} can also be applied to the mixed ices considered in this study. For these ices, the total ice thickness was approximated as being:

\begin{equation}
\label{eq3}
 h_{\textup{total}} = f_1h_1 + f_2h_2 + ...
\end{equation}

\noindent where $f_i$ is the fractional abundance of an ice mixture component and $h_i$ is the thickness that that component would adopt were it a single-component ice. The integrated band strength constants and densities of the ice mixture components used to calculate the compositions and thicknesses of the mixed ices are given in Table~\ref{table1}.

\begin{table}[]
\caption{Densities and IR band strength constants used in this study.}
\label{table1}
\centering
\begin{tabular}{c|ccc}
\hline
Molecule & Density$^a$ & IR Band Position & $A_\nu$$^b$ \\ & (g cm$^{-3}$) & (cm$^{-1}$) & (cm molecule$^{-1}$) \\ \hline
NH$_2$OH    & 1.00              & 1188                    & $9\times10^{-18}$             \\
H$_2$O      & 0.65              & 3300                    & $2\times10^{-16}$             \\
D$_2$O      & 0.74              & 2500                    & $1\times10^{-16}$             \\
CO       & 0.85              & 2139                    & $8.7\times10^{-18}$           \\ \hline
\end{tabular}
\tablefoot{$^a$~The densities of H$_2$O and CO were respectively taken from the works of \citet{dohnalek2003deposition} and \citet{luna2022density}. The density of D$_2$O was obtained by scaling the H$_2$O density by a factor of 20/18.\\
$^b$~The integrated band strength constants of H$_2$O and CO were respectively taken from the works of \citet{Hagen1981} and \citet{gonzalez2022density}. The band strength constant of D$_2$O was obtained by scaling the corresponding value for H$_2$O by 1/2.}
\end{table}

Once the initial column densities and thicknesses of the hydroxylamine-containing ices had been successfully quantified, the ices were exposed to a 15~keV proton beam supplied by the ECRIS that impacted the ice at an angle of 45\textdegree{}. The penetration ranges and linear energy transfers of the proton beams into our ice samples were estimated using the \textit{Stopping and Range of Ions in Matter} (SRIM) software \citep{ziegler2010srim}. In each case, the ice was irradiated in fluence steps of $6\times10^{13}$~protons cm$^{-2}$ after each of which an IR absorption spectrum was acquired so as to discern any radiation-induced changes to the structure and chemical composition of the ice. A summary of the irradiation experiments performed in this study is given in Table~\ref{table2}. 

\begin{table*}[]
\caption{Summary of the irradiation experiments performed in this study. }
\label{table2}
\centering
\begin{tabular}{l|cccccc}
\hline
\multicolumn{1}{c|}{Ice} & \multicolumn{3}{c}{$N$ ($10^{17}$ molecules cm$^{-2}$)} & $h$    & $\rho$ & $F_{\textup{max}}$                \\
\multicolumn{1}{c|}{}    & NH$_2$OH        & H$_2$O or D$_2$O        & CO          & ($\mu$m) & (g cm$^{-3}$)            & ($10^{15}$ protons cm$^{-2}$) \\ \hline
NH$_2$OH                    & 5.37         & -              & -           & 0.295  & 1.00                & 1.20                \\
NH$_2$OH:H$_2$O (1:10)         & 0.87         & 8.75           & -           & 0.450  & 0.68                & 1.86                \\
NH$_2$OH:H$_2$O:CO (1:7:5)     & 1.38         & 9.21           & 7.25        & 0.896  & 0.75                & 1.86                \\
NH$_2$OH:D$_2$O:CO (1:6:3)     & 1.59         & 8.75           & 3.97        & 0.707  & 0.80                & 1.86                \\ \hline
\end{tabular}
\end{table*}

\section{Results}
\label{sec:results}
\subsection{IR Spectroscopic Characterization of Hydroxylamine Ice}
\begin{figure}
    \centering
    \includegraphics[width=1\linewidth]{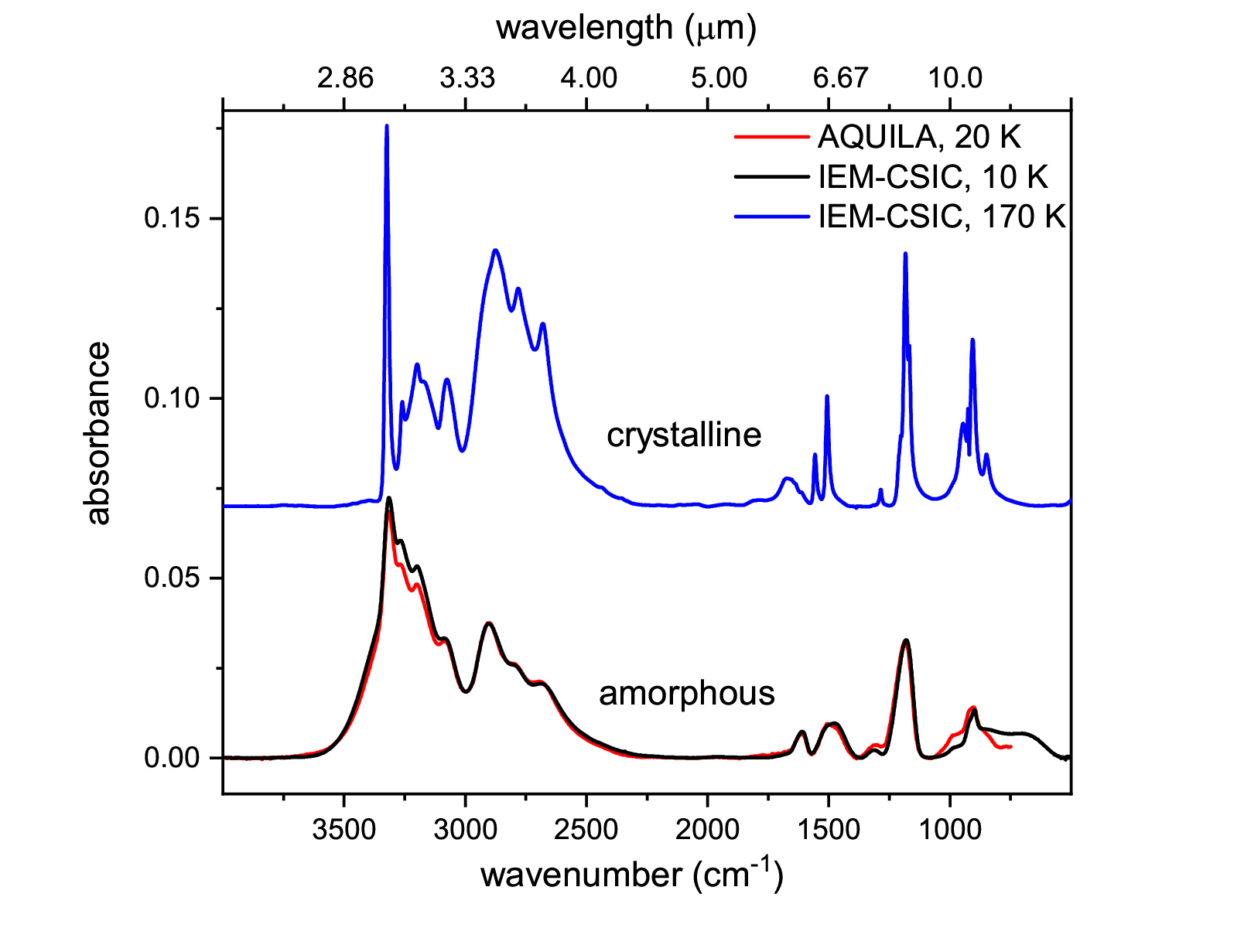}
    \caption{IR absorption spectra of amorphous hydroxylamine ices prepared by vapor deposition at 10~K (set-up at IEM-CSIC; black trace) and 20~K (AQUILA set-up at HUN-REN Atomki; red trace), and of crystalline hydroxylamine ice prepared by depositing at 10~K and warming to 170~K (blue trace). Note that amorphous and crystalline spectra are vertically off-set for clarity.}
    \label{fig1}
\end{figure}

The IR absorption spectra of pure amorphous and crystalline hydroxylamine ice are shown in Fig.~\ref{fig1}. It is worth noting that the spectra of hydroxylamine prepared by vapor deposition at 10~K (IEM-CSIC) and 20~K (Atomki) are nearly identical, and correspond to an amorphous ice structure as is evidenced by the broad absorption features present in these spectra. When the ice prepared at 10~K was warmed at a rate of 10~K min$^{-1}$, an amorphous-to-crystalline phase transition was observed to occur which started at 150~K appeared to be complete by 170~K. Indeed, the IR absorption spectrum acquired at 170~K is characterized by many of the expected features of crystalline ice, including narrower absorption features and band splitting. Further warming of the ice resulted in its sublimation at a temperature of about 180~K.

The major IR absorption features visible in the spectra of amorphous and crystalline hydroxylamine ice are assigned in Table~\ref{table3}. These assignments have been made with reference to the previous works of \citet{nightingale1954vibrational} and \citet{luckhaus1997rovibrational}. To the best of our knowledge, the work of \citet{nightingale1954vibrational} provided the only IR absorption spectra of low-temperature hydroxylamine that were acquired experimentally. Indeed, the method by which they prepared their ices is similar to the vapor deposition methodology used in the present study. The later work of \citet{luckhaus1997rovibrational} was a combined experimental and theoretical study of the ro-vibrational spectrum of gaseous hydroxylamine, with band assignments being made on the basis of calculations. 

We note in passing that, although several previous astrochemical studies have considered the formation of hydroxylamine as a result of, for example, the energetic processing of precursor species \citep{tsegaw2017formation} or the non-energetic hydrogenation of nitric oxide \citep{fedoseev2016simultaneous,nguyen2019experimental}, none of these studies provided a spectrum of neat solid hydroxylamine at low temperatures. The study of \citet{jonusas2016possible} did provide a spectrum of a NH$_2$OH:H$_2$O mixed ice, but this was a consequence of the low vapor pressure of hydroxylamine which precluded those authors from preparing a neat hydroxylamine ice free from contamination from residual water in their high-vacuum set-up.

\begin{figure*}
    \centering
    \includegraphics[width=1\linewidth]{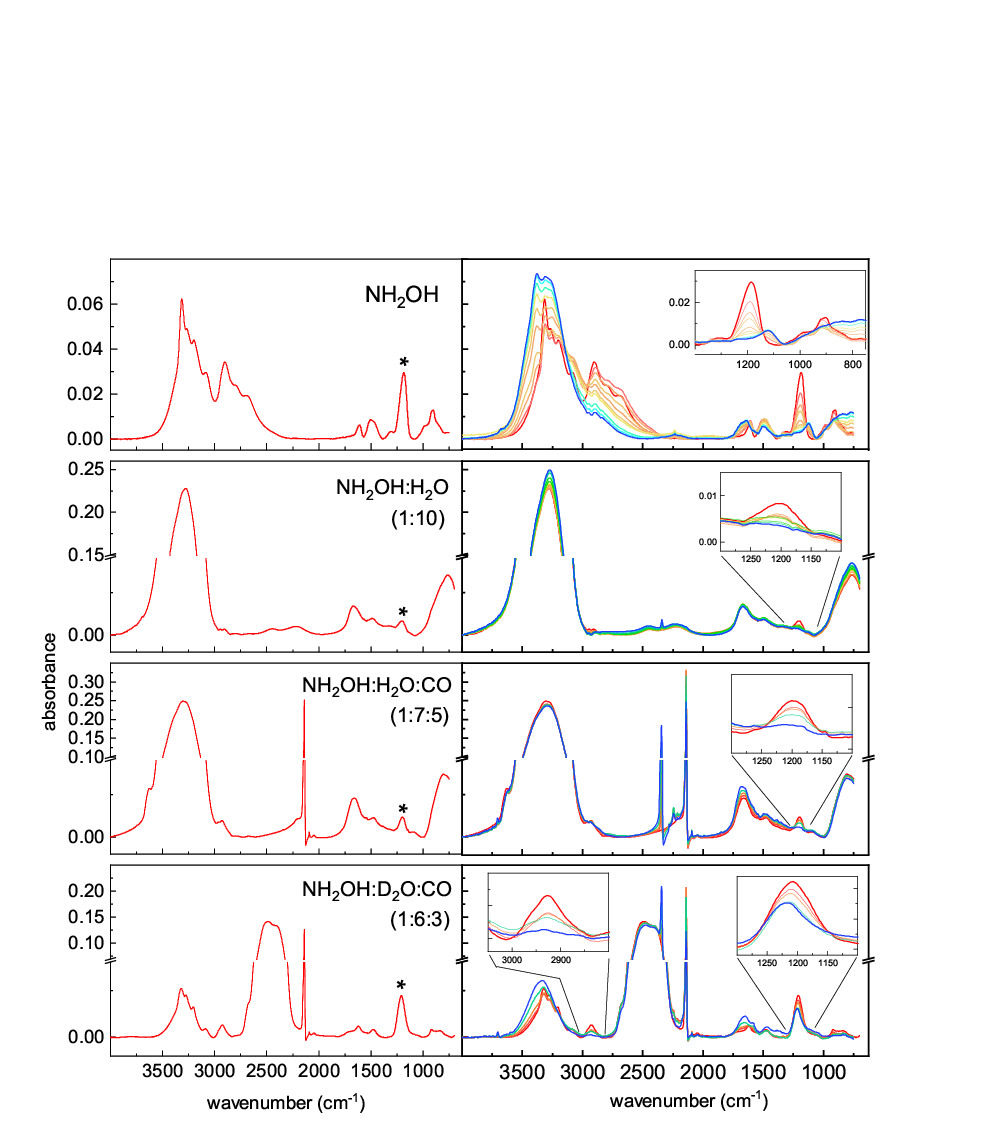}
    \caption{IR absorption spectra of ice mixtures containing hydroxylamine grown by vapor deposition at 20~K using the AQUILA set-up. \textit{Left panels:} IR spectra acquired prior to irradiation, with the absorption band at 1188~cm$^{-1}$ indicated by an asterisk. \textit{Right panels:} Evolution of the IR absorption spectra during irradiation with 15~keV protons. In each panel, the initial IR spectrum is given by the red trace and the final spectrum at the end of the irradiation experiment is given by the blue trace.}
    \label{fig2}
\end{figure*}

IR absorption spectra of different ice mixtures containing hydroxylamine are presented in the left hand-side panels of Fig.~\ref{fig2}. Band assignments for hydroxylamine in these ice mixtures are also provided in Table~\ref{table4}, and it is possible to note that a number of band undergo shifts (typically of less than 10~cm$^{-1}$) in their peak positions, as well as band broadening due to changes in the matrix environment (Fig.~\ref{fig2}, Table~\ref{table4}). Furthermore, a new absorption band at about 1100~cm$^{-1}$ is evident in the spectra of the binary and ternary mixed ices (Fig.~\ref{fig2}), which we tentatively attribute to NH$_3$ \citep{hudson2022ammonia}. We speculate that it is possible that a low-temperature reaction may be occurring in the ice mixture in which hydroxylamine is reduced to ammonia, although further studies are recommended to confirm this. 

\begin{table*}[t]
\caption{Assignments for the IR absorption bands in the spectrum of solid amorphous hydroxylamine ice at 20~K and crystalline hydroxylamine at 170~K.}
\label{table3}
\centering
\begin{tabular}{c|cccccc}
\hline
\multirow{4}{*}{Band Assignment} & \multicolumn{6}{c}{Neat NH$_2$OH Ice}                                                                                            \\  \cline{2-7}
                                 & \multicolumn{3}{c|}{Amorphous Ice}                                     & \multicolumn{3}{c}{Crystalline Ice}                  \\ \cline{2-7} 
                                 & Band Position & Range     & \multicolumn{1}{c|}{A$_\nu$}                    & Band Position & Range     & A$_{\nu\:\textup{(amorph)}}$ / A$_{\nu\:\textup{(cryst)}}$ \\
                                 & (cm$^{-1}$)        & (cm$^{-1}$)    & \multicolumn{1}{c|}{(10$^{-18}$ cm molecule$^{-1}$)} & (cm$^{-1}$)        & (cm$^{-1}$)    &                          \\ \hline
$\nu$ (OH)                           &               &           & \multicolumn{1}{c|}{}                      &               &           &                          \\
$\nu_\textup{asym}$ (NH$_2$)                      & 3314          & 3619-2187 & \multicolumn{1}{c|}{122}                   & 3323          & 3619-2187 & 1.02                     \\
$\nu_\textup{sym}$ (NH$_2$)                       & 3261          &           & \multicolumn{1}{c|}{}                      & 3261          &           &                          \\
                                 & 3199          &           & \multicolumn{1}{c|}{}                      & 3199          &           &                          \\
2$\nu_3$                              & 3080          &           & \multicolumn{1}{c|}{}                      & 3080          &           &                          \\
$\nu_2$, $\nu$ (OH)                       & 2903          &           & \multicolumn{1}{c|}{}                      &               &           &                          \\
                                 &               &           & \multicolumn{1}{c|}{}                      & 2875          &           &                          \\
$\nu_3 + \nu_4$                          & 2789          &           & \multicolumn{1}{c|}{}                      & 2780          &           &                          \\
2$\nu_6$                              & 2682          &           & \multicolumn{1}{c|}{}                      & 2679          &           &                          \\
                                 &               &           & \multicolumn{1}{c|}{}                      &               &           &                          \\ \hline
$\nu_3$, $\delta$ (HNH)                      & 1610          & 1707-1570 & \multicolumn{1}{c|}{1.0}                   & 1670          & 1747-1576 & 2.18                     \\
                                 &               &           & \multicolumn{1}{c|}{}                      & 1558          &           &                          \\ \hline
                                 & 1503          & 1570-1367 & \multicolumn{1}{c|}{3.7}                   & 1507          & 1575-1419 & 0.90                     \\
$\delta$ (HNO)                          & 1473          &           & \multicolumn{1}{c|}{}                      &               &           &                          \\ \hline
tw (NH$_2$)                         & 1313          & 1363-1274 & \multicolumn{1}{c|}{0.3}                   & 1280          & 1334-1245 & 1.03                     \\
                                 &               &           & \multicolumn{1}{c|}{}                      & 1205          &           &                          \\ \hline
w (NH$_2$)                          & 1188          & 1274-1060 & \multicolumn{1}{c|}{9.0}                   & 1183          & 1242-1070 & 1.01                     \\
                                 &               &           & \multicolumn{1}{c|}{}                      & 1168          &           &                          \\
                                 &               &           & \multicolumn{1}{c|}{}                      &               &           &                          \\ \hline
                                 & 987           & 1033-716  & \multicolumn{1}{c|}{6.0}                   & 946           & 1053-716  & 1.80                     \\
$\nu_5$, $\nu$ (NO)                       & 920           &           & \multicolumn{1}{c|}{}                      & 925           &           &                          \\
                                 & 902           &           & \multicolumn{1}{c|}{}                      & 906           &           &                          \\
$\nu_6$, r (NH$_2$)                      & 855           &           & \multicolumn{1}{c|}{}                      & 849           &           &                          \\ \hline
\end{tabular}
\tablefoot{Assignments are made based on the previous works of \citet{nightingale1954vibrational} and \citet{luckhaus1997rovibrational}. Band strength constants for major wavenumber ranges are also given. \\N.B. DB = dangling bonds; asym = asymmetric; sym = symmetric; $\nu$ = stretching, $\delta$ = bending; w = wagging; tw = twisting; r = rocking.}
\end{table*}

\begin{table*}[t]
\caption{Assignments for the IR absorption bands in the spectrum of hydroxylamine in mixtures with H$_2$O (or D$_2$O) and~/~or CO at 20~K. }
\label{table4}
\centering
\begin{tabular}{c|ccc}
\hline
\multirow{2}{*}{Band Assignment} & \multicolumn{3}{c}{NH$_2$OH Band Position in Ice Mixtures (cm$^{-1})$}                                           \\ \cline{2-4} 
                                 & \multicolumn{1}{c|}{NH$_2$OH:H$_2$O (1:10)} & \multicolumn{1}{c|}{NH$_2$OH:H$_2$O:CO (1:7:5)} & NH$_2$OH:D$_2$O:CO (1:6:3) \\ \hline
$\nu$ (OH)                           & \multicolumn{1}{c|}{3698}             & \multicolumn{1}{c|}{3638}                 &                      \\
$\nu_{\textup{asym}}$ (NH$_2$)                      & \multicolumn{1}{c|}{}                 & \multicolumn{1}{c|}{}                     & 3323                 \\
$\nu_{\textup{sym}}$ (NH$_2$)                       & \multicolumn{1}{c|}{}                 & \multicolumn{1}{c|}{}                     & 3272                 \\
                                 & \multicolumn{1}{c|}{}                 & \multicolumn{1}{c|}{}                     & 3204                 \\
2$\nu_3$                              & \multicolumn{1}{c|}{}                 & \multicolumn{1}{c|}{}                     & 3087                 \\
$\nu_2$, $\nu$ (OH)                       & \multicolumn{1}{c|}{2948}             & \multicolumn{1}{c|}{2964}                 & 2969                 \\
                                 & \multicolumn{1}{c|}{2902}             & \multicolumn{1}{c|}{2929}                 & 2929                 \\
$\nu_3 + \nu_4$                          & \multicolumn{1}{c|}{2795}             & \multicolumn{1}{c|}{2792}                 &                      \\
2$\nu_6$                              & \multicolumn{1}{c|}{2678}             & \multicolumn{1}{c|}{2680}                 & 2688                 \\
                                 & \multicolumn{1}{c|}{2458}             & \multicolumn{1}{c|}{}                     &                      \\ \hline
$\nu_3$, $\delta$ (HNH)                      & \multicolumn{1}{c|}{1637}             & \multicolumn{1}{c|}{1649}                 & 1707, 1620           \\
                                 & \multicolumn{1}{c|}{}                 & \multicolumn{1}{c|}{}                     &                      \\ \hline
                                 & \multicolumn{1}{c|}{}                 & \multicolumn{1}{c|}{1544}                 & 1544                 \\
$\delta$ (HNO)                          & \multicolumn{1}{c|}{1487}             & \multicolumn{1}{c|}{1474}                 & 1474                 \\ \hline
tw (NH$_2$)                         & \multicolumn{1}{c|}{1320}             & \multicolumn{1}{c|}{}                     &                      \\
                                 & \multicolumn{1}{c|}{}                 & \multicolumn{1}{c|}{}                     &                      \\ \hline
w (NH$_2$)                          & \multicolumn{1}{c|}{1200}             & \multicolumn{1}{c|}{1197}                 & 1208                 \\
                                 & \multicolumn{1}{c|}{}                 & \multicolumn{1}{c|}{}                     &                      \\
                                 & \multicolumn{1}{c|}{}                 & \multicolumn{1}{c|}{1145}                 &                      \\ \hline
                                 & \multicolumn{1}{c|}{}                 & \multicolumn{1}{c|}{}                     &                      \\
$\nu_5$, $\nu$ (NO)                       & \multicolumn{1}{c|}{918}              & \multicolumn{1}{c|}{}                     & 923                  \\
                                 & \multicolumn{1}{c|}{}                 & \multicolumn{1}{c|}{}                     & 888                  \\
$\nu_6$, r (NH$_2$)                      & \multicolumn{1}{c|}{}                 & \multicolumn{1}{c|}{}                     & 833                  \\ \hline
NH$_3$                              & \multicolumn{1}{c|}{1118}             & \multicolumn{1}{c|}{1085}                 & 1080                 \\ \hline
\end{tabular}
\tablefoot{Assignments are made based on the previous works of \citet{nightingale1954vibrational} and \citet{luckhaus1997rovibrational}. Also included is the band position for the tentatively detected NH$_3$ product observed in the spectra of the ices. \\N.B. DB = dangling bonds; asym = asymmetric; sym = symmetric; $\nu$ = stretching, $\delta$ = bending; w = wagging; tw = twisting; r = rocking.}
\end{table*}

\subsection{Irradiation with 15~keV Protons}
The irradiation of solid-phase hydroxylamine with 15~keV protons results in its destruction. Assuming that this radiolytic destruction follows first-order kinetics, it can be modeled by the equation:

\begin{equation}
\label{eq4}
N = N_0 e^{-\sigma'_{\textup{des}}F}
\end{equation}

\noindent where $F$ is the ion fluence and $\sigma'_{\textup{des}}$ is a parameter termed the \textit{effective} destruction cross-section. It is very important to note that this parameter is not identical to the destruction cross-section term that is typically used in ion-molecule collision physics studies (see the work of \citet{ivlev2023bombardment} as an example). This is primarily due to the fact that true destruction cross-sections can only be defined for collisions in which there is negligible change in the kinetic energy and momentum of the projectile particle, which is not the case in our experiments. Moreover, the effective destruction cross-section given in Eq.~\ref{eq4} takes into account any sputtering of the hydroxylamine into the gas phase as a result of the collision of the projectile proton, which is not the case for true destruction cross-sections that only take into account molecular destruction through radiolytic dissociation. Nevertheless, the effective destruction cross-section parameter is useful to gauge the radiolytic resistance of molecules to ionizing radiation used in laboratory experiments to mimic the effects of galactic cosmic rays (see, for example, previous works by \citet{mate2011infrared,mate2018stability,mate20212,herrero2022stability}) and thus can be utilized to gain insights into astrochemical processes.

As shown in Eq.~\ref{eq1}, the molecular column density of hydroxylamine at a given ion fluence is proportional to the intensity of its absorption bands as measured from acquired IR absorption spectra. Substituting Eq.~\ref{eq1} into Eq.~\ref{eq4} yields:

\begin{equation}
\label{eq5}
\textup{ln}\left(\frac{I_F}{I_0}\right)=-\sigma'_\textup{des}F
\end{equation}

\noindent where $I_0$ and $I_F$ respectively represent the integrated area of a characteristic hydroxylamine absorption band prior to irradiation and at some given fluence $F$.

It is also important to note that, as the hydroxylamine-containing ice is irradiated, its chemical composition will change and new products can be formed which may take part in secondary reactions. This may lead to a deviation from the exponential decay of hydroxylamine described by Eq.~\ref{eq4}; however, it is possible to assume that this equation is accurate during the initial stages of irradiation when radiolytic dissociation should be the only major contributor to hydroxylamine destruction. 

The right hand-side panels of Fig.~\ref{fig2} depict the evolution of the IR absorption spectra of the four ices containing hydroxylamine that were considered in this study as a function of proton fluence. Although it would be ideal to calculate effective destruction cross-sections from all the apparent hydroxylamine absorption bands (as these different cross-sections will provide information on the various possible radiolytic mechanisms taking place), these spectra demonstrate that this is not possible for several of the absorption bands due either to their overlap with other components in the ice (i.e., H$_2$O or CO), or with the absorption bands of products formed as a result of radiolysis. For instance, the hydroxylamine NH and OH stretching modes in the 3000~cm$^{-1}$ region overlap with the broad and strong OH stretching modes of water. Furthermore, the hydroxylamine HNH and HNO bending modes in the $1500-1450$~cm$^{-1}$ region of the spectrum overlap with the broad bending mode of water at about 1600~cm$^{-1}$, as well as with a number of features attributable to radiolytic products. The hydroxylamine NO stretching mode at about 920~cm$^{-1}$ also overlaps with the libration mode of water.

Fortunately, the hydroxylamine NH$_2$ wagging mode at about 1188~cm$^{-1}$ remains fairly isolated from other absorption bands throughout the irradiation experiment, and is therefore a good candidate to measure with the aim of quantifying effective destruction cross-sections. It is to be noted, however, that in the ice mixture containing D$_2$O, the NH$_2$ wagging mode of hydroxylamine overlaps with the D$_2$O bending mode. To overcome this problem, we have assumed that D$_2$O is not destroyed to any significant degree during irradiation and therefore a constant contribution of D$_2$O to this absorption band at 1188~cm$^{-1}$ could be subtracted at different fluences to determine the extent of hydroxylamine destruction. However, the hydroxylamine OH and NO stretching modes respectively located at about 3000~cm$^{-1}$ and 920~cm$^{-1}$ are free from overlap with other absorption features in the ice mixture containing D$_2$O, and thus could be utilized to quantify the effective destruction cross-section. 

\begin{figure*}
    \centering
    \includegraphics[width=1\linewidth]{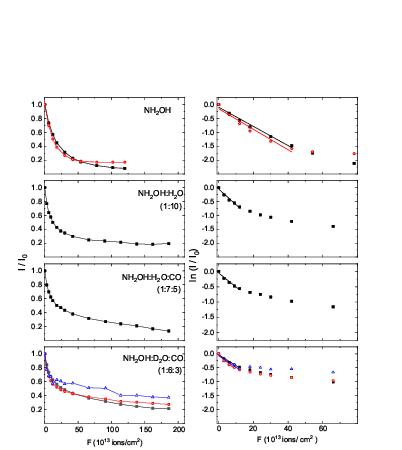}
    \caption{Hydroxylamine IR band decay as a function of proton fluence. \textit{Left panels:} Normalized absorption band area decay for three distinct IR vibrational modes (black squares: NH$_2$ wagging mode; red circles: OH stretching mode; blue triangles: NO stretching mode). Note that lines between the individual data points are not fits and are plotted solely to guide the eye. \textit{Right panels:} Natural logarithmic representation of the data plotted in the left hand-side panels. Note that, in this case, the straight lines are linear trendlines fitted to data points at low proton fluences.}
    \label{fig3}
\end{figure*}

Overall, the radiolysis-induced decay of the hydroxylamine NH$_2$ wagging mode at approximately 1188~cm$^{-1}$ was used to quantify the effective destruction cross-section of hydroxylamine in all four studied ices. Additionally, the decay of the OH stretching mode at 2600~cm$^{-1}$ was used to quantify the effective destruction cross-section in the neat hydroxylamine ice as well as the ternary mixture containing D$_2$O. The effective destruction cross-section of hydroxylamine in this latter ice mixture was also calculated by measuring the decay of the NO stretching mode at about 920~cm$^{-1}$. The left hand-side panels of Fig.~\ref{fig3} show the normalized decay trends of the relevant hydroxylamine absorption band areas. Effective destruction cross-sections have been determined by fitting a linear trend to the natural logarithmic representation of this data (this is shown in the right hand-side panels of Fig.~\ref{fig3}) and only considering the initial fluence steps of the irradiation process.

Having evaluated the effective destruction cross-section, it is possible to estimate the half-life dose, $D_{1/2}$; this is defined as the amount of energy that must be deposited into the ice in order to reduce the abundance of hydroxylamine to half of its original value. From Eq.~\ref{eq4}, the fluence required to achieve this reduction in hydroxylamine is given as:

\begin{equation}
\label{eq6}
F_{1/2}=\frac{\textup{ln}\:(2)}{\sigma'_{\textup{des}}}
\end{equation}

\noindent The half-life dose can therefore be expressed as:

\begin{equation}
\label{eq7}
D_{1/2}=\frac{F_{1/2}f_{\textup{E}}E}{N_0} =\frac{\textup{ln}\:(2)f_{\textup{E}}E}{\sigma'_{\textup{des}}N_0}
\end{equation}

\noindent where $E$ is the kinetic energy of the impacting protons (i.e., 15~keV) and $f_{\textup{E}}$ is the fraction of this energy deposited into the ice during irradiation. If it is assumed that the energy imparted by the ions to the ice sample is done so in a homogeneous manner, then Eq.~\ref{eq7} can be re-written in terms of the linear energy transfer (LET) of the protons and the density of the ice:

\begin{equation}
\label{eq8}
D_{1/2}=\frac{\textup{ln}\:(2)\:\times\:\textup{LET}}{\rho\sigma'_{\textup{des}}}
\end{equation}

\noindent The LET of the protons, together with the density of the ice and the effective destruction cross-section, may be used to quantify another parameter often used in radiation chemistry studies: the so-called \textit{G}-value of hydroxylamine. This is defined as the number of hydroxylamine molecules that undergo radiolysis per 100~eV of energy that is deposited into the ice and is given as:

\begin{equation}
\label{eq9}
G=100\:\frac{\rho\sigma'_{\textup{des}}}{\textup{LET}}
\end{equation}

\noindent As stated previously, the linear energy transfers of the 15~keV protons in the four ices considered in this study were calculated using the SRIM software. The calculated linear energy transfers, effective destruction cross-sections, and \textit{G}-values are summarized in Table~\ref{table5}.

\begin{table*}[t]
\caption{Summary of the calculated effective destruction cross-sections, linear energy transfers, half-life doses, and \textit{G}-values of hydroxylamine-containing ices irradiated by 15~keV protons.}
\label{table5}
\centering
\begin{tabular}{l|cccccc}
\hline
\multirow{2}{*}{Ice} & \multicolumn{3}{c}{$\sigma'_{\textup{des}}$ ($10^{15}$~cm$^2$)} & LET      & $D_{1/2}$          & $G$                    \\
                     & 1188~cm$^{-1}$    & 2600~cm$^{-1}$   & 920~cm$^{-1}$    & keV mm$^{-1}$ & eV molecule$^{-1}$ & molecules per 100~eV \\ \hline
NH$_2$OH                & 3.5 ± 0.2    & 3.7 ± 0.4   &             & 56       & 6.6 ± 0.5     & 11 ± 1               \\
NH$_2$OH:H$_2$O (1:10)     & 5.6 ± 0.6    &             &             & 23       & 1.5 ± 0.5     & 46 ± 5               \\
NH$_2$OH:H$_2$O:CO (1:7:5) & 4.6 ± 0.5    &             &             & 36       & 2.8 ± 0.5     & 25 ± 2               \\
NH$_2$OH:D$_2$O:CO (1:6:3) & 3.9 ± 0.4    & 5.4 ± 0.9   & 4.2 ± 0.7   & 38       & 2.9 ± 0.5     & 24 ± 2               \\ \hline
\end{tabular}
\end{table*}

\subsection{Radiolytic Chemistry at 20~K}
\begin{figure}
    \centering
    \includegraphics[width=1\linewidth]{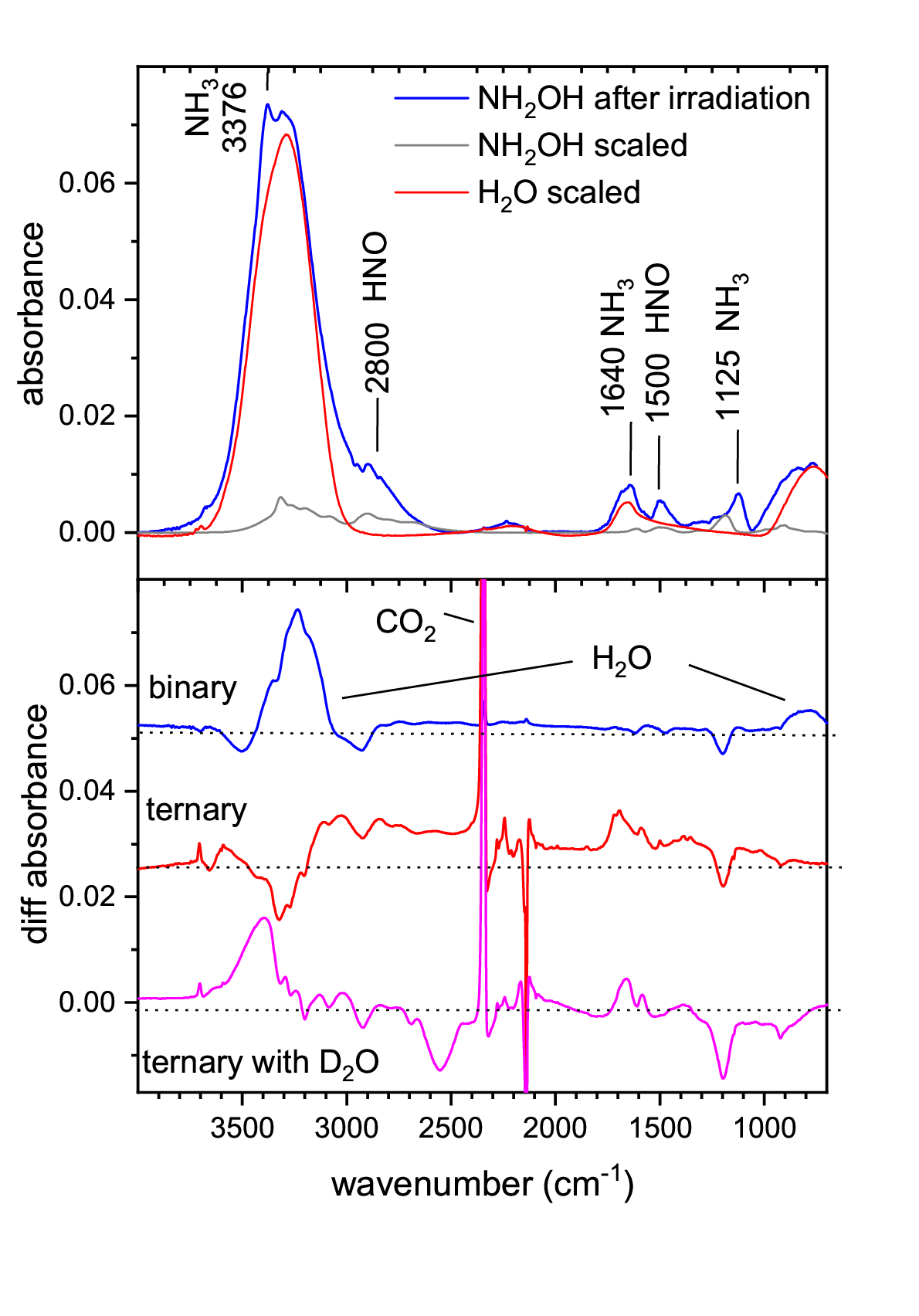}
    \caption{IR spectra acquired to assess the major radiolytic products of the studied ices. \textit{Top panel:} Spectrum of neat hydroxylamine ice after irradiation with 15~keV protons (blue trace), together with spectra of amorphous water (red trace) and neat hydroxylamine ices (gray trace) scaled to match the amount of water formed and the hydroxylamine that survives at the end of the irradiation. \textit{Bottom panel:} Difference spectra (i.e., the initial, pre-irradiation spectrum subtracted from that acquired at the end of the experiment) of the three ice mixtures containing hydroxylamine that were considered in this study. Note that two of the spectra in this panel have been vertically off-set for clarity.}
    \label{fig4}
\end{figure}

The irradiation of neat hydroxylamine ice with 15~keV protons resulted in the almost complete destruction of this molecule, as indicated by the data plotted in Fig.~\ref{fig3}. To better identify IR absorption features attributable to radiolytic product species, the spectrum acquired at the end of the irradiation experiment has been plotted alongside the scaled spectra of neat hydroxylamine and neat water ices at 20~K (Fig.~\ref{fig4}), thus allowing for the maximum amount of surviving hydroxylamine to be visualized. It is evident from Fig.~\ref{fig4} that water is a major radiolytic product, although absorption features that can be ascribed to other molecules are also evident. These include bands at 3376, 1640, and 1125~cm$^{-1}$, whose positions are most commensurate with the asymmetric stretching, degenerated deformation, and symmetric deformation vibrational modes of ammonia, respectively. This is in agreement with the previous theoretical work of \citet{wang2010thermal}, who proposed that the major dissociation reaction is:

\begin{equation}
\label{eq10}
    \ce{2NH2OH -> H2O + NH3 + HNO}
\end{equation}

\noindent Indeed, absorption bands visible at about 2800 and 1500~cm$^{-1}$ (Fig.~\ref{fig4}) are likely associated with the NH stretching and HNO bengind modes of the HNO molecule, respectively \citep{jacox1973matrix}.

In the case of the irradiation of the NH$_2$OH:H$_2$O (1:10) mixed ice, it is difficult to observe the formation of new bands due to the intense and broad absorption features of water. Indeed, when analyzing the difference spectrum (i.e., the spectrum obtained when subtracting the initial, pre-irradiation spectrum from that acquired after the end of the irradiation process), it is only possible to recognize IR absorption features associated with the formation of water (Fig.~\ref{fig4}). It is, however, possible to note that the radiolytic destruction of hydroxylamine is still quite extensive, as evidenced by the decay of the band at 1188~cm$^{-1}$ with increasing proton fluence (Fig.~\ref{fig3}). As such, the dominant chemical process taking place is likely similar to that suggested by \citet{feller2003nonparametrized}:

\begin{equation}
\label{eq11}
    \ce{2NH2OH + 2H2O -> 2H2O + 2NH3 + O2}
\end{equation}

Perhaps unsurprisingly, the richest chemistry in our experiments was observed during the irradiation of the ternary NH$_2$OH:H$_2$O(D$_2$O):CO ice mixtures (Fig.~\ref{fig4}). No attempt at characterizing the radiolytic products has been made in the case of the experiment involving D$_2$O, primarily due to a lack of information as to the wavenumber positions of the absorption bands of the relevant deuterated molecules in the solid phase. 

\begin{figure*}
    \centering
    \includegraphics[width=1\linewidth]{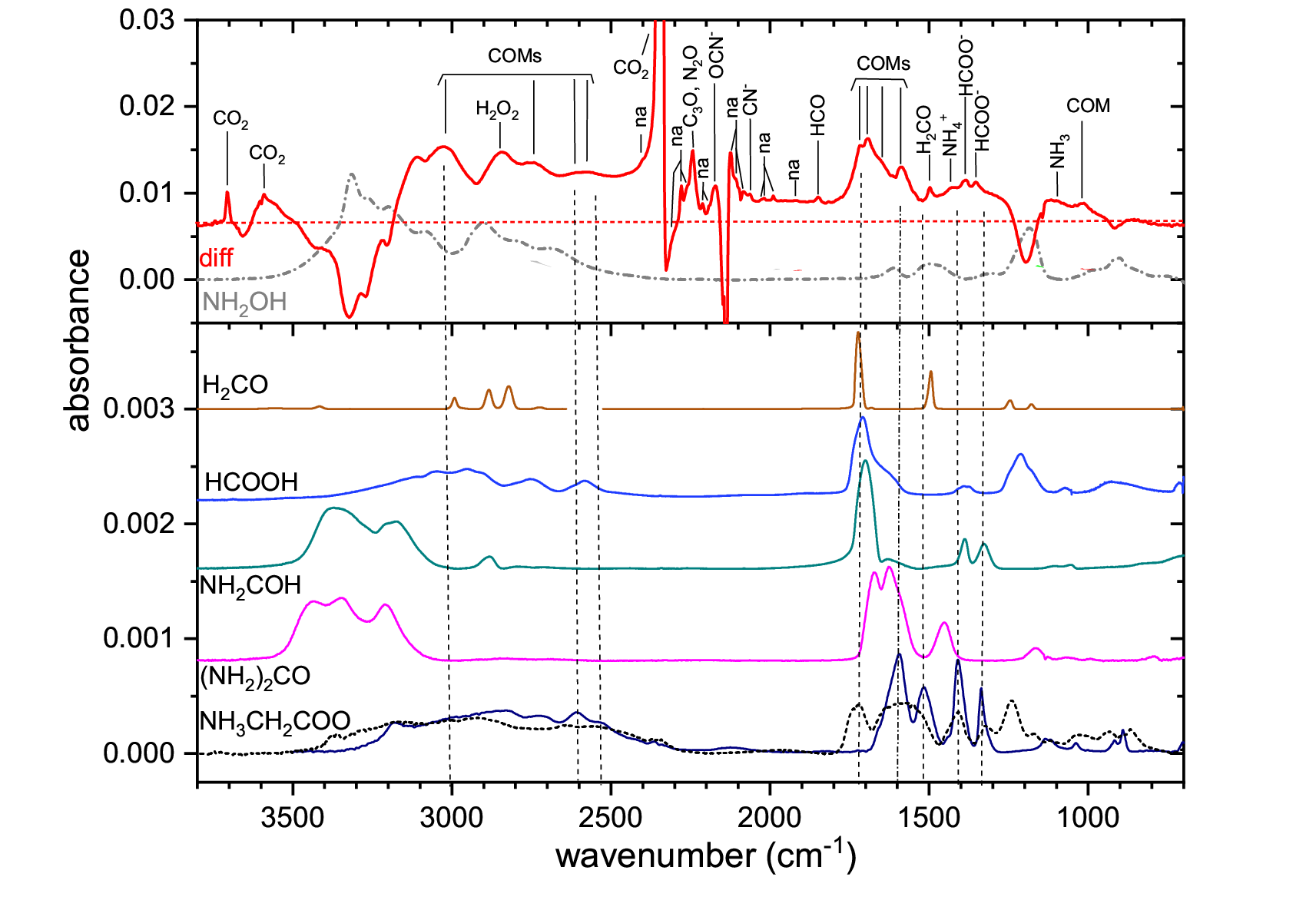}
    \caption{Analysis of the potential COMs formed as a result of irradiating hydroxylamine-containing ices. \textit{Top panel:} Difference spectrum (i.e., the initial, pre-irradiation spectrum subtracted from the final spectrum acquired after the end of the irradiation process) of the ternary NH$_2$OH:H$_2$O:CO ice (red solid trace). Also shown is the spectrum of neat hydroxylamine (gray dashed trace) to compare against the negative peaks associated with molecular destruction in the difference spectrum. \textit{Bottom panel:} IR absorption spectra of candidate complex organic molecular products assembled from the literature \citep{gerakines1996ultraviolet,brucato2006infrared,mate2011infrared,bergantini2014processing,timon2021infrared}. From top to bottom, these are spectra of: formaldehyde, formic acid, formamide, urea, and neutral (dashed line) and zwitterionic (solid line) glycine.}
    \label{fig5}
\end{figure*}

The same difference spectrum given in the lower panel of Fig.~\ref{fig4} for the irradiated NH$_2$OH:H$_2$O:CO ice mixture is plotted in the top panel of Fig.~\ref{fig5}, where it is possible to more clearly identify the positive absorption bands of a number of radiolytic product molecules. The strongest of these absorption bands is that at 2343~cm$^{-1}$ which is ascribed to the asymmetric stretching mode of CO$_2$; combination modes for this molecule are also present at 3700 and 3591~cm$^{-1}$. Other simple molecules detected by means of this spectrum include: H$_2$O$_2$ (2843~cm$^{-1}$), C$_3$O and~/~or N$_2$O ($2244-2238$~cm$^{-1}$), HNCO (2263~cm$^{-1}$), OCN$^-$ (2174~cm$^{-1}$), CN$^-$ (2065~cm$^{-1}$), HCO (1850~cm$^{-1}$), and NH$_3$ (1096~cm$^{-1}$; very weak). These assignments were made on the basis of the previous works of \citet{gerakines1996ultraviolet}, \citet{brucato2006infrared}, \citet{loeffler2006synthesis}, and \citet{galvez2010ammonium}. Furthermore, features associated with NH$_4^+$ (1431~cm$^{-1}$) and HCOO$^-$ (1386~cm$^{-1}$) ions are observable in the difference spectrum (Fig.~\ref{fig5}). A list of band assignments for these radiolytic products is given in Table~\ref{table6}.

To assist with the identification of more complex organic species such as formic acid (HCOOH), formamide (NH$_2$CHO), or urea, the low-temperature spectra of these molecules in the solid phase are depicted in Fig.~\ref{fig5}. These spectra have been taken from the works of \citet{brucato2006infrared}, \citet{gerakines1996ultraviolet}, \citet{bergantini2014processing}, and \citet{timon2021infrared}, and correspond to neat ices prepared at temperatures of $15-20$~K. Considering that changes in the matrix environment of a cryogenic solid typically alter the positions of its main IR absorption bands by about 10~cm$^{-1}$, it is possible to propose that these three complex organic molecules are good radiolytic product candidates. The presence of both carbamic acid (NH$_2$COOH), whose main IR absorption features are at 1691, 1451, and 1320~cm$^{-1}$, as well as carbonic acid (H$_2$CO$_3$) which displays three equally intense features at 1714, 1598, and 1313~cm$^{-1}$, also cannot be excluded \citep{moore1991infrared,brucato1997carbonic,james2020systematic,james2021systematic,marks2023thermal}. IR absorption spectra of the neutral and zwitterionic forms of glycine, taken from the work of \citet{mate2011infrared}, are also depicted in Fig.~\ref{fig5} and it is possible to note that the main absorption features of this amino acid coincide with regions of the difference spectrum in which there is an absorbance gain. As such, it is possible to suggest that glycine is among the radiolytic products.

\begin{table}[t]
\caption{Assignments for the new bands attributed to product formation during the irradiation of the NH$_2$OH:H$_2$O:CO ternary ice mixture using 15~keV protons. }
\label{table6}
\centering
\begin{tabular}{cc}
\hline
Band Peak (cm$^{-1}$) & Assignment \\ \hline
3700             & CO$_2$        \\
3591             & CO$_2$        \\
3021 (broad)       & COM        \\
2843 (broad)       & H$_2$O$_2$       \\
2741 (broad)       & COM        \\
2610 (broad)       & COM        \\
2575 (broad)       & COM        \\
2400 (weak)      & na         \\
2343             & CO$_2$        \\
2306 (weak)      & na         \\
2280             & na         \\
2260 (weak)      & na         \\
2244             & C$_3$O or N$_2$O \\
2212             & na         \\
2192 (weak)      & na         \\
2174             & OCN$^-$       \\
2124             & na         \\
2106 (weak)      & na         \\
2062             & CN$^-$        \\
2028             & na         \\
2018             & na         \\
1990             & na         \\
1920 (weak)      & na         \\
1850             & HCO        \\
1717             & COM        \\
1693             & COM        \\
1650 (broad)       & COM        \\
1588             & COM        \\
1498             & H$_2$CO       \\
1431             & NH$_4^+$       \\
1384             & HCOO$^-$      \\
1350             & HCOO$^-$      \\
1096 (weak)      & NH$_3$        \\
1014             & COM        \\ \hline
\end{tabular}
\tablefoot{Assignments have been made largely on the previous works of \citet{hudson1999laboratory}, \citet{gerakines1996ultraviolet}, \citet{brucato2006infrared}, \citet{loeffler2006synthesis}, and \citet{galvez2010ammonium}. Tentative assignments for complex organic molecules (COMs) are given in Table~\ref{table7}. \\N.B. na = band not assigned.}
\end{table}

\begin{table*}[t]
\caption{Possible complex organic molecules present in the irradiated NH$_2$OH:H$_2$O:CO ternary ice based on the observed IR absorption features detailed in Table~\ref{table6} and Figs.~\ref{fig5} and~\ref{fig6}.}
\label{table7}
\centering
\begin{tabular}{l|ccc}
\hline
Molecule                & Characteristic IR Absorption Bands (cm$^{-1}$)          & Reference                & Detected? \\ \hline
Carbamic Acid (H$_2$NCOOH) & 3462, 3140 (br), 1691, 1451, 1320                  & \citet{bossa2008carbamic}      & Possible             \\
Carbonic Acid (H$_2$CO$_3$)   & 2850 (w), 2630, 1714 (=), 1508 (=), 1313 (=), 1040 & \citet{ioppolo2021vacuum}    & No                   \\
Formamide (H$_2$NCHO)      & 3368-3181, 2881, 1708 (s), 1631 (s), 1388, 1328    & \citet{brucato2006infrared}    & Likely               \\
Formic Acid (HCOOH)     & 1685, 1380 (w), 1211 (m)                           & \citet{bergantini2014processing} & Yes                  \\
Glycine (H$_2$NCH$_2$COOH)    & 3500-3200, 1725, 1650, 1423, 1326, 1250            & \citet{mate2011infrared}       & Possible             \\
Urea (H$_2$NCONH$_2$)         & 1672, 1626, 1594, 1454                             & \citet{timon2021infrared}      & Likely               \\ \hline
\end{tabular}
\tablefoot{N.B. (w) = weak band; (m) = medium-strength band; (s) = strong band; (br) = broad band; (=) = bands are of approximately equal intensity.}
\end{table*}

\begin{figure}
    \centering
    \includegraphics[width=1\linewidth]{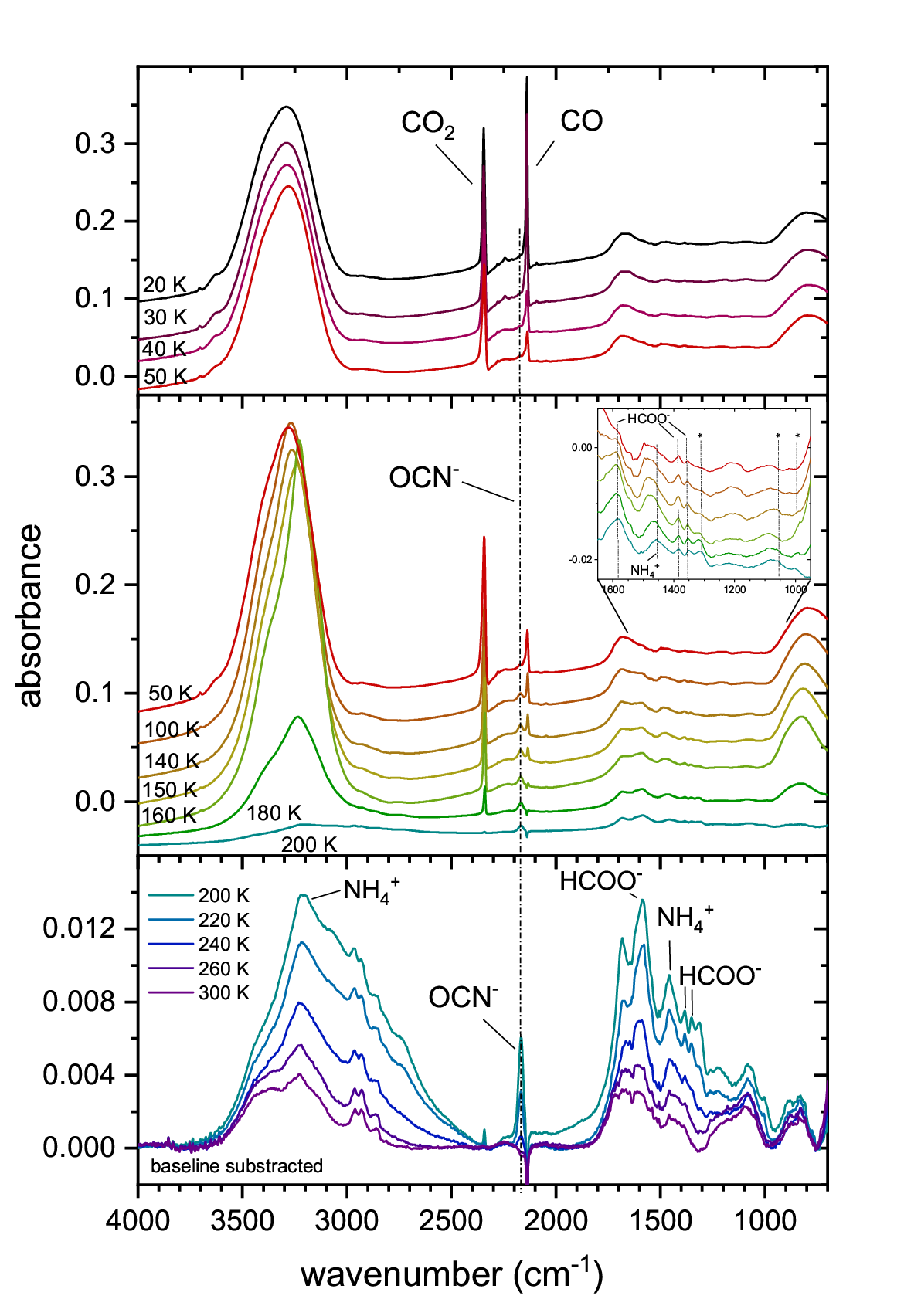}
    \caption{Evolution of the IR absorption spectra of the 15~keV proton-irradiated NH$_2$OH:H$_2$O:CO ice upon warming at a rate of 1~K min$^{-1}$.}
    \label{fig6}
\end{figure}

To obtain additional information on the radiation chemistry taking place in the ternary NH$_2$OH:H$_2$O:CO ice as a result of its exposure to 15~keV protons, a post-irradiative temperature-programmed desorption (TPD) experiment was performed in which the ice was warmed at a rate of 1~K min$^{-1}$. IR absorption spectra acquired during this TPD experiment are depicted in Fig.~\ref{fig6}. Sublimation of the components of the irradiated ice mixture is anticipated to follow a trend of decreasing volatility and, indeed, CO was the first molecule observed to sublime, followed by CO$_2$. The sublimation of CO$_2$ was accompanied by the greater formation of OCN$^-$ (i.e., in addition to that produced radiolytically), whose absorption band peak shifts from 2174~cm$^{-1}$ to 2165~cm$^{-1}$. This band begins to decrease during the sublimation of water and eventually disappears by a temperature of 220~K. 

An interesting result of the TPD experiment was that, during the phase transition of water from amorphous to crystalline at $140-160$~K \citep{Hagen1981}, the absorption bands associated with the HCOO$^-$ (1588, 1386, and 1350~cm$^{-1}$) and NH$_4^+$ (1450~cm$^{-1}$) ions were observed to grow (Fig.~\ref{fig6}). It is possible that the rearrangement of water molecules in the ice matrix allowed for the better diffusion of these ions to yield inorganic salts, including ammonium formate (NH$_4^+$HCOO$^-$) and ammonium cyanate (NH$_4^+$OCN$^-$). Similar results were previously reported by \citet{galvez2010ammonium} and \citet{mate2012cyanate}. Once the sublimation of water was practically completed at 200~K, the IR absorption spectrum was dominated by the features of these two inorganic salts: bands associated with NH$_4^+$ are clearly visible at 3200 and 1450~cm$^{-1}$ in the spectrum acquired at 200~K, together with bands associated with OCN$^-$ at 2175~cm$^{-1}$ and with HCOO$^-$ at 1588, 1386, and 1350~cm$^{-1}$. Other bands also visibly increase in intensity upon warming to 200~K, such as those at 1312, 1060, and 1000~cm$^{-1}$ (marked with asterisks in the middle panel insert of Fig.~\ref{fig6}), whose carriers could not be unambiguously identified.

\begin{figure}
    \centering
    \includegraphics[width=1\linewidth]{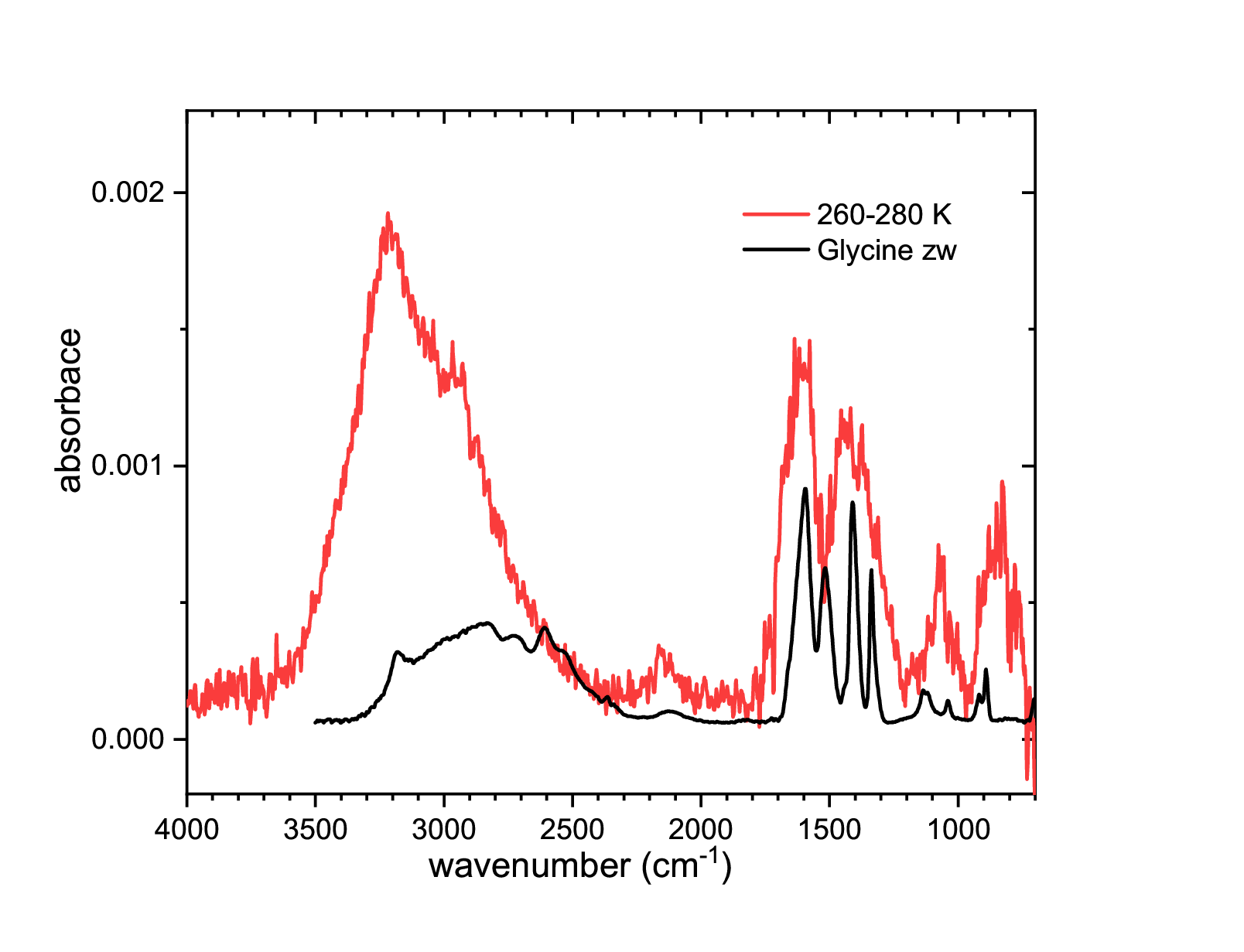}
    \caption{Difference spectrum acquired by using IR absorption spectra acquired at temperatures of 260 and 280~K during the TPD experiment (red trace) shown alongside a spectrum of neat zwitterionic glycine (black trace).}
    \label{fig7}
\end{figure}

At temperatures greater than 200~K, ammonium formate was observed to thermally decompose into ammonia and formic acid, which in turn undergo sublimation. Ammonium cyanate is also removed from the deposition substrate, but at a slightly higher temperature of 220~K, as discussed previously. This allows for other, more refractory organic species present in lower abundances to be identified in higher temperature IR absorption spectra. To aid in constraining the identity of these species, it is possible to compute the difference spectrum across a given temperature interval. For a difference spectrum produced as a result of subtracting the spectrum acquired at a temperature $T_1$ from that acquired at another temperature $T_2$ (i.e., $T_2-T_1$, where $T_1>T_2$), it is possible to associate positive absorption bands in the resultant difference spectrum with material that has undergone sublimation across this temperature interval. Fig.~\ref{fig7} depicts the difference spectrum obtained in this way by using IR absorption spectra acquired at 260 and 280~K during the TPD experiment, along with the spectrum of neat zwitterionic glycine for comparative purposes. Given the similarities between these spectra, together with previous reports of glycine undergoing sublimation at 280~K under ultrahigh-vacuum conditions \citep{esmaili2018glycine}, it is possible that glycine is indeed a product of the radiolysis of the ternary ice mixture considered in this study. 

\section{Discussion}
\label{sec:discussion}
\subsection{IR Absorption Spectroscopy}
One of the main aims of this study was the characterization of the IR absorption spectra of hydroxylamine-containing ices under conditions relevant to astrochemistry. In this regard, our experiments have demonstrated that the preparation of hydroxylamine ice by background condensation of the gas at $10-20$~K results in an amorphous structure. Warming the resultant ice to higher temperatures does not result in any thermal decomposition; however, a phase transition to a more ordered crystalline form begins at 150~K and is largely complete by 170~K (Fig.~\ref{fig1}). The formation of this crystalline phase was evident through the appearance of new, resolved IR features as well as the sharpening of bands, as has been observed for several other molecular ices \citep{hudson2014infrared,hudson2017infrared,hudson2020infrared,timon2021infrared,hudson2022infrared,mifsud2023systematic,mifsud2024systematic}: for instance, a single broad band at about 1503~cm$^{-1}$ in the spectrum of the amorphous ice is resolved to a doublet structure with peaks at 1558 and 1507~cm$^{-1}$ in the spectrum of the crystalline phase. Further warming of the hydroxylamine ice to temperatures of 180~K and greater resulted in the sublimation of the ice, consistent with the view of hydroxylamine being a less volatile species than water under conditions relevant to astrochemistry \citep{molpeceres2023processing}.

The strongest absorption features in the IR spectrum of solid hydroxylamine are the OH and NH stretching modes. However, these bands overlap with the low-frequency wing of the broad and strong OH stretching mode of water, and thus are not easy to identify in ice mixtures dominated by water. As such, it is likely that these two hydroxylamine vibrational modes are not good candidates for the search for icy hydroxylamine in space environments using high-resolution and high-sensitivity instruments such as the \textit{James Webb Space Telescope}. Instead, we propose that the band at about 1188~cm$^{-1}$, which remains clearly identifiable in ice mixtures containing both H$_2$O and CO (Fig.~\ref{fig2}), is a better candidate for this purpose.  

\subsection{Radiolytic Destruction of Hydroxylamine}
Another goal of our study was to quantify the radio-resistance of solid hydroxylamine during its irradiation by 15~keV protons which were used as analogues of galactic cosmic rays. To achieve this goal, effective destruction cross-sections were quantified for all the investigated hydroxylamine-containing ices by following the decay of the IR absorption band at about 1188~cm$^{-1}$ (Table~\ref{table5}). This band corresponds to the NH$_2$ wagging mode and possibly samples the dissociation reaction: 

\begin{equation}
\label{eq12}
    \ce{NH2OH} \curly \ce{HNOH + H}
\end{equation}

Analysis of the effective destruction cross-sections given in Table~\ref{table5} reveals a slightly slower rate of radiolytic destruction for neat hydroxylamine ice compared to the binary mixture with H$_2$O and the ternary mixture with H$_2$O and CO. Moreover, since the effective destruction cross-section of hydroxylamine in the ternary ice mixture containing D$_2$O could be calculated not only from the decay of the NH$_2$ wagging mode at 1188~cm$^{-1}$ but also from the decay of the OH and NH stretching modes at about 2600 and 920~cm$^{-1}$ respectively, it is possible to compare these values to one another. It is evident that, to a first approximation, the effective destruction cross-sections calculated from these IR absorption bands are similar within experimental error.

It is also possible to compare the calculated half-life doses shown in Table~\ref{table5} to those previously calculated for other complex organic molecules relevant to astrochemistry, such as glycine, urea, methyl isocyanate, and 2-aminooxazole \citep{mate2011infrared,mate2018stability,mate20212,herrero2022stability}. Comparisons reveal that the half-life dose of hydroxylamine is significantly lower than those calculated for these other organic species of interest (see \citet{mate20212} and references therein for more information), implying that hydroxylamine in interstellar ices is significantly less radio-resistant during cosmic ray irradiation. This result, taken with the fact that the half-life doses of complex organic molecules are generally less than those of the underlying dust grains in interstellar clouds \citep{mate2016high,molpeceres2017structure} as well as the arguments previously made by \citet{molpeceres2023processing} as to the overall low abundance of hydroxylamine in the gas-phase, helps to explain why this molecule has thus far only been detected in one interstellar source \citep{rivilla2020prebiotic}. However, it is important to emphasize that the efficient radiolytic destruction of hydroxylamine in interstellar ices likely contributes to a rich solid-phase chemistry leading to the formation of complex organic molecules relevant to biology, as discussed below. 

\subsection{Radiolytic Chemistry}
Our final objective was to elucidate the radiolytic chemistry of hydroxylamine in astrophysical ice analogues. The postulated importance of hydroxylamine in the synthesis of prebiotic molecules provided a strong motivation to study its radiation chemistry not only as a single-component ice, but also in more realistic interstellar ice analogues containing H$_2$O and CO. Our experiments have demonstrated the unambiguous detection of a number of simple product molecules, including H$_2$O, NH$_3$, H$_2$O$_2$, and HNO (Fig.~\ref{fig2}, Table~\ref{table6}), in all considered ices. The formation of these molecules can be rationalized through the following reactions following the dissociation of hydroxylamine and the recombination of the resultant radical species:

\begin{align} 
\label{eq:set1}
        \ce{NH2OH} \curly \ce{NH2 + OH}\\
        \ce{NH2OH} \curly \ce{H2NO + H &-> HNO + 2H} \\
        \ce{NH2OH} \curly \ce{HNOH + H &-> HNO + 2H} \\
        \ce{NH2 + H &-> NH3} \\
        \ce{OH + H &-> H2O} \\
        \ce{2OH &-> H2O2}
\end{align}

Of course, other reaction routes may also potentially contribute to the synthesis of these simple product molecules. It is interesting to note, however, that the radiolysis of hydroxylamine is analogous to that of isoelectronic molecules such as CH$_3$OH or H$_2$O$_2$ \citep{hudson2001radiation,loeffler2006synthesis,bennett2007mechanistical}. As an example, the hydroxylamine reactions described by Eqs.~14-16 are analogous to the formation of H$_2$CO and CH$_4$ during the radiolysis of CH$_3$OH.

Our experiments have also demonstrated that the 15~keV proton irradiation of hydroxylamine-containing ices results in the production of complex organic molecules. Although it is likely that a myriad of chemical reactions could conceivably contribute to the formation of such products, it is reasonable to assume that the initial reaction must involve CO, since this is the only carbon-bearing molecule initially included in our ice mixtures. As such, CO may react with excited or suprathermal radical species; particularly those that are produced in abundance as a result of the radiolytic dissociation of the parent hydroxylamine molecule and are also more likely to be mobile within the ice matrix, such as OH, NH$_2$, and hydrogen atoms. Therefore, the most likely products to form as a result of the reaction between CO and these radicals are:

\begin{align} 
\label{eq:set2}
        \ce{CO + OH &-> HOCO} \\
        \ce{CO + NH2 &-> NH2CO} \\
        \ce{CO + H &-> HCO}
\end{align}

Each of these products (i.e., HOCO, NH$_2$CO, and HCO) is an important starting point for a new network of reactions which results in the formation of new complex organic molecules, some of which may be prebiotic in nature. Each of these possible networks is discussed in further detail in the subsections below. 

\subsubsection{HOCO-Initiated Chemistry}
Following the synthesis of HOCO within the irradiated ice mixture, it is possible that this species itself may undergo reactions with the more mobile abundant radicals within the ice (i.e., OH, NH$_2$, and hydrogen atoms). These reactions respectively lead to the synthesis of carbonic acid, carbamic acid, and formic acid:

\begin{align} 
\label{eq:set3}
        \ce{HOCO + OH &-> H2CO3} \\
        \ce{HOCO + NH2 &-> H2NCOOH} \\
        \ce{HOCO + H &-> HCOOH}
\end{align}

Of these three acid species, we have only been able to definitively detect formic acid among the radiolytic products of our experiment (Fig.~\ref{fig5} and Table~\ref{table7}). Interestingly, our study has also demonstrated that formic acid may be retained in the solid phase at temperatures greater than 200~K (i.e., above its anticipated sublimation temperature; \citet{molpeceres2022hydrogen}) as the inorganic ammonium formate salt. It should be noted, however, that \citet{molpeceres2022hydrogen} calculated binding energies for isolated formic acid molecules on amorphous solid water ice as proxies for sublimation temperatures, which is not quite representative of our present experiments. Nevertheless, the observation of ammonium formate at higher temperatures is of particular importance, as recent studies have suggested that ammonium salts could account for an important reservoir of nitrogen in icy Solar System environments \citep{poch2020ammonium} and, therefore, could conceivably associate with deprotonated organic acids under conditions relevant to astrochemistry.

Conversely, the detection of the carbonic and carbamic acids remains more elusive due to the complexity of the IR spectra acquired during and after the irradiation process. However, we are able to speculate on the likelihood of their formation: starting with carbonic acid, which was recently detected in the interstellar medium \citep{sanz2023discovery}, we suggest that its presence within our irradiated ices is unlikely due to mismatches between the observed IR absorption bands in our spectra and those reported in the literature for carbonic acid \citep{ioppolo2021vacuum}. However, we note that the absence of carbonic acid in our experiments cannot be unequivocally linked to its non-formation, due to the known ease with which this molecule undergoes dissociation upon exposure to ionizing radiation \citep{ioppolo2021vacuum}. As such, it is possible that carbonic acid did form in our experiments, but was quickly dissociated to yield other molecules such as H$_2$O and CO$_2$. In the case of carbamic acid, although there is a somewhat better agreement between the IR absorption features in our spectra and those reported in the literature for this molecule \citep{james2020systematic,james2021systematic,marks2023thermal}, it is difficult to definitively confirm its presence due to the possibility that other organic species that could conceivably form in the irradiated ice may contribute to at least some of the observed IR bands.

\subsubsection{NH$_2$CO-Initiated Chemistry}
The reaction of NH$_2$CO with OH radicals, NH$_2$ radicals, and hydrogen atoms could result in the direct synthesis of carbamic acid, urea, and formamide, respectively: 

\begin{align} 
\label{eq:set4}
        \ce{NH2CO + OH &-> H2NCOOH} \\
        \ce{NH2CO + NH2 &-> H2NCONH2} \\
        \ce{NH2CO + H &-> NH2CHO}
\end{align}

As described previously, the presence of carbamic acid in our irradiated ice cannot be definitively confirmed. However, based on the reasonably good agreement between band peak positions in our IR absorption spectra and those in the literature \citep{sivaraman2013infrared}, formamide is more than likely present within our irradiated ice (Table~\ref{table7}). Moreover, several features in the $1800-1600$~cm$^{-1}$ region of the spectrum also show good agreement with the expected absorption bands of urea \citep{timon2021infrared}, tentatively suggesting the presence of this molecule too. However, other bands in our IR spectra suggested to be associated with urea (such as that at about 1500~cm$^{-1}$) appear to be shifted to slightly higher wavenumbers compared to their literature positions and, as such, the detection of urea should be considered to be tentative.

\subsubsection{HCO-Initiated Chemistry}
The reaction of HCO with OH radicals, NH$_2$ radicals, and hydrogen atoms could lead to the formation of formic acid, formamide, and formaldehyde, respectively: 

\begin{align} 
\label{eq:set4}
        \ce{HCO + OH &-> HCOOH} \\
        \ce{HCO + NH2 &-> NH2CHO} \\
        \ce{HCO + H &-> H2CO}
\end{align}

The chemistry leading to the formation of formic acid and formamide has already been discussed. In addition, we note the unambiguous detection of formaldehyde in our experiments (Fig.~\ref{fig5} and Table~\ref{table6}). Therefore, all products of the reactions of HCO with radicals formed as a result of the radiolytic dissociation of the hydroxylamine parent species were detected in the irradiated ternary ice mixture. 

\subsubsection{Other Reactions}
Our IR absorption spectra also evidence other types of reactions taking place within the irradiated ice mixture. Among these is the reaction between ground-state CO and excited CO, which contributes to the formation of CO$_2$ \citep{jamieson2006understanding,ivlev2023bombardment}. It should be noted that non-energetic reactions, such as those involving HOCO intermediates, could also contribute in a non-negligible manner to the presence of CO$_2$ in the irradiated ice mixture \citep{ioppolo2011surface,Qasim2019,Molpeceres2023CO2}, although the exact mechanism of the reaction is still under debate and will be the subject of forthcoming work. Other routes toward the formation of CO$_2$ are also possible, including those making use of radiolytic products within the ice. For instance, the reaction between CO and $^3$O may be possible due to the injection of energy by the impinging projectile ions which may be used to overcome the energy barrier associated with this spin-forbidden reaction \citep{Minissale2013}.

Acid-base chemistry also takes place in the ice, as is evidenced by the formation of OCN$^-$ \citep{hudson2001formation,van2004quantitative}. In this scenario, a precursor molecule (likely isocyanic acid, HNCO) donates a proton to either water or ammonia to yield a salt product:

\begin{align} 
\label{eq:set5}
        \ce{H2O + HNCO &-> H3O+ + OCN-} \\
        \ce{NH3 + HNCO &-> NH4+ + OCN-}
\end{align}

\noindent Indeed, IR spectral signatures of ammonium cyanate were obsered in spectra acquired during the TPD experiment at temperatures below 220~K (Fig.~\ref{fig6}).

One product which was not observed in our irradiation experiments was hydrazine (NH$_2$NH$_2$), despite the fact that the synthesis of this molecule as a result of the combination of two NH$_2$ radicals seems relatively straightforward. The reason for this is not fully understood; however, we speculate that it may be related to the relatively low abundance of ammonia in our ices (as evidenced by the weak spectral features of that molecule; Table~\ref{table6}). To the best of our knowledge, the synthesis of hydrazine as a result of the irradiation of astrophysical ice analogues has only been reported in those ices that were initially rich in ammonia \citep{zheng2008formation,henderson2015direct,shulenberger2019electron}. Therefore, given that ammonia was never a dominant species in the ices considered in this study, it is entirely possible that the radiolytic formation of hydrazine either occurred to an extent below the limits of spectroscopic detection or, alternatively, was precluded entirely. 

At this point, we must emphasize that our analysis of the radiolytic chemistry occurring within the investigated ices likely only represents a small portion of the chemical processes taking place. Indeed, our analysis primarily focuses on radical-radical combination reactions and the role of excited neutral molecules. However, it is likely that ion-based chemistry plays an equally significant role in explaining the observed chemical complexity. For example, the observed formation of salts is challenging to fully rationalize without invoking chemical processes that involve ions. Despite its importance, ion-based chemistry is less straightforward to describe, and many specific reaction mechanisms are poorly understood. As such, conclusions formed regarding reactions involving ionic species would likely be speculative at best, and we therefore consider these reactions to be beyond the scope of the present work. Nonetheless, we emphasize the importance of these species in the chemistry occurring in our irradiated ices and plan to explore this subject more fully in future work.

\subsubsection{Proposed Radiolytic Synthesis of Glycine}
Finally, and in light of the chemical connection between hydroxylamine and glycine \citep{snow2007gas}, we also searched for the latter species among the products of the irradiated NH$_2$OH:H$_2$O:CO ternary ice mixture. However, such a detection is challenging if based solely on IR absorption spectroscopy, and we are only able to tentatively claim a detection in our experiments. This is primarily due to the generally weak absorption features that were tentatively attributed to glycine in our spectra, compounded by the complexity of the possible co-existence of neutral and zwitterionic glycine depending on the polarity of the local matrix environment. However, it is to be noted that the large degree of similarity between the IR absorption spectrum of zwitterionic glycine and that of the final residue after our TPD experiment (Fig.~\ref{fig7}), combined with the disappearance of these absorption features at the expected desorption temperature of glycine, constitutes an optimistic outlook for the presence of glycine as a radiolytic product in our study. 

\section{Summary and Conclusions}
\label{sec:conclusions}
In this study, we have considered the IR absorption spectra of neat and mixed hydroxylamine astrophysical ice analogues prepared through vapor condensation onto a cooled substrate. Furthermore, we have studied the radio-resistance of hydroxylamine in these ices when irradiated using 15~keV protons, as well as the ensuing chemistry. Our main findings can be summarized as follows:

\begin{enumerate}
    \item The vapor deposition of hydroxylamine onto a cooled substrate at $10-20$~K results in an amorphous ice which crystallizes upon heating to temperatures greater than 150~K in a phase change process that is largely complete by 170~K. Hydroxylamine ice sublimes from the substrate upon further heating to 180~K.
    \item IR absorption band strengths of amorphous hydroxylamine have been estimated from the amount of water formed after its complete radiolytic destruction (see appendix for more information).
    \item Our spectroscopic characterizations have determined that the NH$_2$ wagging mode of hydroxylamine at about 1188~cm$^{-1}$ is the most appropriate band by which this species could be detected in icy environments in space, and that is has an estimated band strength constant of $A_\nu=9\times10^{-18}$~cm molecule$^{-1}$. We strongly emphasize that this value is merely an estimate, and that dedicated experiments are required to accurately measure the band strength constants of this molecule.
    \item Effective destruction cross-sections of hydroxylamine irradiated by 15~keV protons at 20~K have been determined for the neat ice, the binary mixture with H$_2$O, and the ternary mixture with H$_2$O and CO; from which half-life doses were computed. These half-life doses are about one order of magnitude lower than those of other commonly studied complex organic molecules such as glycine and urea, indicating the low radio-resistance of solid hydroxylamine in space environments. This may make its future detection in icy interstellar grains challenging. 
    \item The radiolytic chemistry of hydroxylamine in a ternary ice mixture also containing H$_2$O and CO is rich and produces a number of organic species including formaldehyde, formic acid, and urea. Inorganic ammonium salts (i.e., ammonium formate and ammonium cyante) were also formed. Other species of varying complexity, such as carbamic acid, formamide, and glycine have been tentatively detected as radiolytic products; although more work is required to confirm a definitive detection for these species. 
    \item Potential reaction routes and networks have been suggested to rationalize the diversity of the products detected in our experiments.
    \\
\end{enumerate}

\begin{acknowledgements}
The authors gratefully acknowledge support from the Europlanet RI through the Transnational Access Project Grant no. 22-EPN3-074. The Europlanet RI has received funding from the European Union's Horizon 2020 Research Innovation Programme under grant agreement no. 871149. This study is also based on work from the COST Action CA20129 MultIChem, supported by COST (European Cooperation in Science and Technology). Belén Maté, Ramón J. Peláez, and Juan Ortigoso acknowledge support provided by grant no. PID2020-113084GB-I00 funded by MICIU/AEI/10.13039/501100011033. Germán Molpeceres acknowledges the support provided by the grant no. RYC2022-035442-I funded by MICIU/AEI/10.13039/501100011033 and ESF+. Germán Molpeceres also acknowledges support received from the project 20245AT016 (Proyectos Intramurales CSIC). Víctor M. Rivilla acknowledges the Spanish State Research Agency (AEI) through project no. PID2022-136814NB-100 from the Spanish Ministy of Science, Innovation, and Univeristies~/~State Agency of Research MICIU/AEI/10.13039/501100011033 and by "ERDF: A Way of Making Europe". Víctor M. Rivilla also acknowledges support from the grant no. RYC2020-029387-1 funded by MICIU/AEI/10.13039/501100011033 and by "ESF: Investing in Your Future", as well as from the Consejo Superior de Investigaciones Científicas (CSIC) and the Centro de Astrobiología (CAB) through the project no. 20225AT015 (Proyectos Intramurales Especiales del CSIC). Víctor M. Rivilla is grateful for support from the grant CNS2023-144464 funded by MICIU/AEI/10.13039/501100011033 and by "European Union NextGenerationEU~/~PRTR". Sergio Ioppolo thanks the Danish National Research Foundation for support through the Centre of Excellence 'InterCat' (grant agreement no. DNRF150). Zoltán Juhász is grateful for support from the Hungarian Academy of Sciences through the János Bolyai Research Scholarship. 
\end{acknowledgements}

\bibliographystyle{aa} 
\bibliography{bibliography.bib}
\balance{}

\begin{appendix}
\label{appendixA}
\section{Hydroxylamine IR Band Strength Estimation}
Our experiment on the irradiation of neat hydroxylamine ice at 20~K led to practically the complete destruction of this molecule, as can be seen in Fig.~\ref{fig3}. As discussed previously, hydroxylamine is anticipated to primarily decompose as per the reaction:

\begin{equation}
\label{app.eq1}
\ce{2NH2OH \curly H2O + NH3 + HNO}
\end{equation}

Indeed, the formation of water is evident in the top panel of Fig.~\ref{fig4}, which illustrates an IR absorption spectrum of amorphous water scaled to match the amount of water formed radiolytically. Therefore, by measuring the column density of water from this scaled spectrum, it is possible to make use of the stoichiometric relationship given in Eq.~\ref{app.eq1} to calculate the molecular column density of hydroxylamine molecules which must dissociate to yield the measured abundance of water (this was done using the band strength constant of water indicated in Table~\ref{table1}). Due to the very extensive radiolytic destruction of hydroxylamine in this experiment, this calculated column density of hydroxylamine is assumed to be equal to the initial abundance present. 

Having estimated the initial column density of hydroxylamine present in the neat ice prior to irradiation, $N_0$, and knowing the initial integrated absorption of the IR band at 1188~cm$^{-1}$, $I_0$, it is possible to rearrange Eq.~\ref{eq1} to calculate the band strength constant:

\begin{equation}
\label{app.eq2}
A_\nu = \textup{ln}(10)\:\frac{I_0}{N_0}
\end{equation}

In principle, it is also possible to use this methodology to calculate the initial column density of hydroxylamine by making use of the umbrella vibrational mode of ammonia at about 1125~cm$^{-1}$. However, the strength constant of this band is about one order of magnitude weaker than that of the OH stretching modes of water and thus the uncertainties associated with measuring the integrated absorption of the ammonia band are larger. Moreover, previous works have indicated that the band strength constant associated with the umbrella vibrational mode of ammonia varies significantly depending on the chemical nature of the ice \citep{kerkhof1999infrared}. As such, the use of the IR absorption features of ammonia to estimate the initial column density of hydroxylamine was not pursued. 

Of course, this estimation method is associated with a number of sources of error, such as the uncertainty in the estimation of H$_2$O formation due to the IR absorption bands of this molecule overlapping with those of NH$_3$ and HNO (see top panel of Fig.~\ref{fig4}). Another source of error is the assumption that the dissociation mechanism shown in Eq.~\ref{app.eq1} is the only reaction by which hydroxylamine is destroyed during irradiation. In reality, it is possible that other reaction mechanisms that do not hold the same stoichiometry as Eq.~\ref{app.eq1} may contribute to the decay of hydroxylamine, as well as ion impact-induced sputtering. However, in the absence of dedicated studies aimed at quantifying the IR absorption band strength constants of hydroxylamine, the present method of estimation should be considered an acceptable substitute. 

\end{appendix}
\end{document}